\documentclass[amssymb, preprintnumbers, showpacs, showkeys,aps,prb,superscriptaddress,twocolumn]{revtex4-1}

\usepackage{color} 
\usepackage{xcolor} 
\usepackage{hyperref}
\usepackage{slashed}
\usepackage{graphicx}
\usepackage{amsmath}
\usepackage{latexsym}
\usepackage{epstopdf}
\usepackage{amsmath}
\usepackage{enumitem}
\usepackage{amssymb,amsmath}
\usepackage{multirow}
\usepackage{mathtools}
\usepackage{bbm}
\usepackage{bm}
\usepackage{amsfonts}
\usepackage{amssymb}
\usepackage{calligra}
\usepackage{calrsfs}
\usepackage{makecell}
\usepackage[normalem]{ulem}
\usepackage[USenglish,american]{babel}

\expandafter\ifx\csname package@font\endcsname\relax\else
 \expandafter\expandafter
 \expandafter\usepackage
 \expandafter\expandafter
 \expandafter{\csname package@font\endcsname}%
\fi
\begin{document}

\title{Exploring quantum quasicrystal patterns: a variational study}

\author{A. Mendoza-Coto}%
\email{alejandro.mendoza@ufsc.br}
\affiliation{Departamento de F\'\i sica, Universidade Federal de Santa Catarina, 88040-900 Florian\'opolis, Brazil}%

\author{R. Turcati}%
\affiliation{Departamento de F\'\i sica, Universidade Federal de Santa Catarina, 88040-900 Florian\'opolis, Brazil}%

\author{V. Zampronio}%
\affiliation{Departamento  de  F\'isica  Te\'orica  e  Experimental,  Universidade  Federal  do  Rio  Grande  do  Norte}%

\author{R. D\'iaz-M\'endez}%
\affiliation{Ericsson B. A. Digital Services R\&D, Ericsson Building 8, 164 40 Kista, Sweden}
\affiliation{Department of Physics, KTH Royal Institute of Technology, 106 91 Stockholm, Sweden}%

\author{T. Macr\`i}%
\affiliation{Departamento  de  F\'isica  Te\'orica  e  Experimental,  Universidade  Federal  do  Rio  Grande  do  Norte and  International  Institute  of  Physics,  Natal-RN,  Brazil}%

\author{F. Cinti}%
\email{fabio.cinti@unifi.it}
\affiliation{Dipartimento di Fisica e Astronomia, Universit\`a di Firenze, I-50019, Sesto Fiorentino (FI), Italy}
\affiliation{INFN, Sezione di Firenze, I-50019, Sesto Fiorentino (FI), Italy}
\affiliation{Department of Physics, University of Johannesburg, P.O. Box 524, Auckland Park 2006, South Africa}

\begin{abstract}
	
	We study the emergence of quasicrystal configurations produced purely by quantum fluctuations in the ground-state phase diagram of interacting bosonic systems. 
    By using a variational mean-field approach, we determine the relevant features of the pair interaction potential that stabilize such quasicrystalline states in two dimensions.
    Unlike their classical counterpart, in which the interplay between only two wave vectors determines the resulting symmetries of the solutions, the quantum picture relates in a more complex way to the instabilities of the excitation spectrum.           
    Moreover, the quantum quasicrystal patterns are found to emerge as the ground state with no need of moderate thermal fluctuations.  
    The study extends to the exploration of the excitation properties and the possible existence of super-quasicrystals, i.e. supersolid-like quasicrystalline states in which the long-range non-periodic density profile coexist with a non-zero superfluid fraction. 
    Our calculations show that, in an intermediate region between the homogeneous superfluid and the normal quasicrystal phases, these exotic states indeed exist at zero temperature. Comparison with full numerical simulations provides a solid verification of the variational approach adopted in this work.
\end{abstract}

\maketitle

\section{Introduction}
The exploration of patterns with peculiar symmetries like stripe phases, smectic liquid crystals, cluster crystals and quasicrystals is a leading research direction in many-body physics, unveiling a large amount of fascinating phenomena in soft matter\cite{Zhou2019,Zhang2021}, superconductivity\cite{Mller,Fratini2010}, nonlinear optical systems \cite{PhysRevE.66.046220,PhysRevLett.82.4627,PhysRevLett.74.258} and long-range interacting systems in general \cite{Bttcher2021,PhysRevX.6.041039,Tanzi2021,PhysRevA.96.013627,PhysRevLett.119.215302,Deguchi2012,Sato2017,Defenu21}. In this context quasicrystals are one of the most intriguing examples, as particles self-assemble in a long-range ordered pattern which is at the same time non-periodic, thus been able to exhibit forbidden crystalline ordering like five-, ten- and twelve-fold rotational symmetry in $2$D.

For classical systems it has been shown that quasicrystalline phases may be originated thanks to the interplay between two specific length-scales in the interaction potential of particle ensembles~\cite{PhysRevLett.126.158001,Dotera2011,PhysRevE.70.021202,Kumar2017}. Many studies, based on mean-field and molecular dynamics approaches, have actually observed the stabilization of decagonal and dodecagonal cluster quasicrystals in soft macromolecular systems at finite temperatures by using this type of interactions~\cite{Barkan2011,Barkan2014,likos01}. While a recent theoretical work have surprisingly revealed the stability of those structures also at zero temperature for a particular case~\cite{Dotera2014}, the extent to which classical cluster quasicrystals can be stable in the absence of thermal fluctuations is a matter of debate. 

On the other hand, quantum cluster quasicrystals have been studied by imposing external quasi-periodic potentials to bosonic systems, so creating quasicrystalline structures in, e.g., two-dimensional optical lattices~\cite{Viebahn2019,Sbroscia2020}. Interestingly, the competition of interactions and quasiperiodicity generate a wide range of significant phases, such as supersolidity and Bose glasses~\cite{PhysRevLett.126.110401,ciardi2021finite,Cinti2010,cinti14,deabreu2020superstripes,PhysRevLett.120.060407,PhysRevLett.123.210604}. 

In the absence of external potentials and for small
temperatures, superfluidity was also investigated in a model relevant to quantum cluster quasicrystal~\cite{Pupillo2020,Fa2019}. 
By using quantum Monte Carlo approaches, it was found that moderate quantum fluctuations make dodecagonal structures to persist, leading to a small but finite local superfluid phase.
Yet, by increasing fluctuations, a structural transition from quasicrystal to cluster triangular crystal takes place.
In this scenario a natural question to ask is whether it is possible to produce cluster quasicrystal phases solely as a joined effect of quantum fluctuations and a properly designed interaction potential between particles. 
To our knowledge, the stabilization of self-assembled cluster quasicrystal states at zero temperature is a completely open question for generic interactions.

\begin{figure*}[t]
\includegraphics[width=\textwidth]{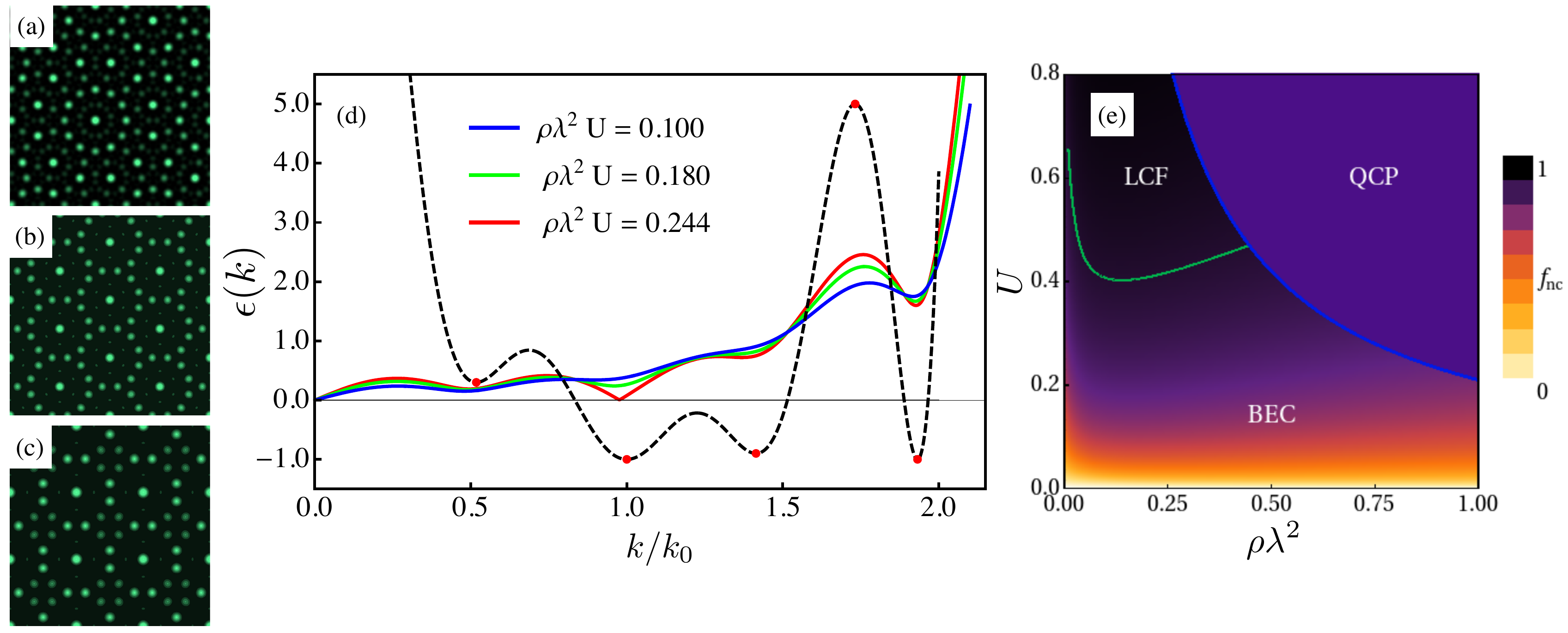}
\caption{ (a-c) Real space density for representative ansatze resulting from the minimization of the energy of Eq.~(\ref{Hdim}) for an ensemble of bosons interacting via the potential of Eq.~(\ref{pot}). (a) periodic pattern with twelve fold dodecagonal symmetry, (b) periodic pattern with six fold hexagonal symmetry, and (c) periodic pattern with four fold square symmetry. The parameters of the potential are: $\hat{v}(0)=20$, $\hat{v}(\sqrt{2}k_0)=5$, $\hat{v}(\sqrt{3}k_0)=-0.9$, $\hat{v}(\sqrt{2-\sqrt{3}}k_0)=0.3$, $\sigma=1.2$ and $\rho\lambda^2U=1.0$ (see Section~\ref{singlemode}). (d) Bogoliubov excitation spectrum $\epsilon(k)$ in units of $\hbar^2/(m\lambda^2)$ for different values of the product $\rho\lambda^2U$ for the model considered in Sec.~(\ref{Pdiag}), selecting $\hat{v}(\sqrt{2-\sqrt{3}}k_0)=0.3$ ($k_0=1$). The Fourier transform of the potential $\hat{v}(k)$ for the same parameters is also plotted (black dashed curve) for comparison with the energy excitation spectrum. (e) Phase diagram and non-condensed fraction $f_{nc}$ computed self-consistently from Eq.~(\ref{eq:cond_frac}). The green curve separates the homogeneous low condensation regime ($f_{nc}>0.9$) from the region in which the condensed fraction is significant. The blue curve separates the homogeneous from the dodecagonal cluster quasicrystal phase. The color scale corresponds to the non-condensed fraction only within the homogeneous phase.}
\label{fig1t}
\end{figure*}

In this paper we address this problem through a variational mean-field approach (VMF) \cite{prestipino18,prestipino2018}. Our study allows to identify the ingredients of the pair interaction potential needed for the stabilization of cluster quasicrystal states at zero temperature.
As a result we present a systematic study of the mean field phase diagrams 
for a class of bosonic models displaying quasicrystalline phases as well as other more common periodic and homogeneous superfluid phases. 
As an illustrative example, Fig.~\ref{fig1t} depicts some stable density profiles for pattern configurations with twelve-fold dodecagonal symmetry (Fig.~\ref{fig1t}a), six-fold hexagonal symmetry (Fig.~\ref{fig1t}b) and four-fold square symmetry (Fig.~\ref{fig1t}c) decorated with twelve smaller cluster.
Here we denote as triangular crystal phase (TCP) and square crystal phase (SqCP) to any phase with the hexagonal and square symmetries respectively, independently of the detailed structure of the unitary cell of the pattern.     
We highlight that the methodology developed in the present work is general and it can be used to design different types of potentials capable to stabilize other modulated patterns.
Additionally, we also probe the stability of the quasicrystal phase against generic perturbations of the pair interaction potential, showing that such a phase is not a result of a fine adjustment of the form of the potential.

Finally we investigate the low energy excitations in the homogeneous phase. The main result illustrated in Fig.~\ref{fig1t}d shows the Bogoliubov excitation spectrum for different values of $\rho\lambda^2U$ properly tuning quasicrystal potentials, whereas Fig.~\ref{fig1t}e illustrates the phase diagram for the Bose-Einstein condensate fraction. The existence of superfluidity within quasicrystalline phases is also discussed. Our findings indicate the existence of supersolid-like quasicrystallline states in an intermediate phase between the homogeneous superfluid phase and a normal cluster quasicrystal phase.
Results obtained from our variational calculations are compared with its analogs obtained from the numerical integration of the Gross-Pitaevskii equation showing an excellent agreement.

The paper is organized as follows: Section~\ref{Method} introduces the microscopic model specifying the methodology that was applied. 
Section~\ref{singlemode} aims to present some important considerations by means of a first single mode approximation. 
A characterisation of a phase diagram varying the interaction strength and fixing the parametric potentials is proposed in Section~\ref{phasediagram}. Furthermore, in Section~\ref{stability}, we inspect the stability of the quantum quasicrystal phase in its ground state. Within the Bogoliubov approach, Section~\ref{excitations} discusses the excitation properties of the system. Section~\ref{supersolid} examines the chance to observe supersolid features in the system. Finally, Section~\ref{conclusions} delivers some conclusions and remarks.

\section{Model and methodology}\label{Method}

We examine an ensemble of interacting bosons confined in a $2$D with mass $m$ and position $\mathbf{q}_i$. 
The dynamics is described by the Hamiltonian
\begin{equation}
 \hat{H}=-\frac{\hbar^2}{2m}\sum_i \nabla_i^2+V_0\sum_{i<j}v(\vert\mathbf{q}_i-\mathbf{q}_j\vert),
 \label{Hdim}
\end{equation}
where $V_0$ is the interaction strength of the two-body potential $v(r)$. We scale Eq.~\eqref{Hdim} by introducing a characteristic length $\lambda$ and energy $\epsilon_0=\hbar^2/(m\lambda^2)$ so that the dimensionless term $U=V_0m\lambda^2/\hbar^2$ controls the zero-temperature physics at a fixed density $\rho\lambda^2$. Likewise, the dimensionless single-particle coordinate results as $\mathbf{r}_i=\mathbf{q}_i/\lambda$.

\begin{figure}[t!]
	\includegraphics[width=\linewidth]{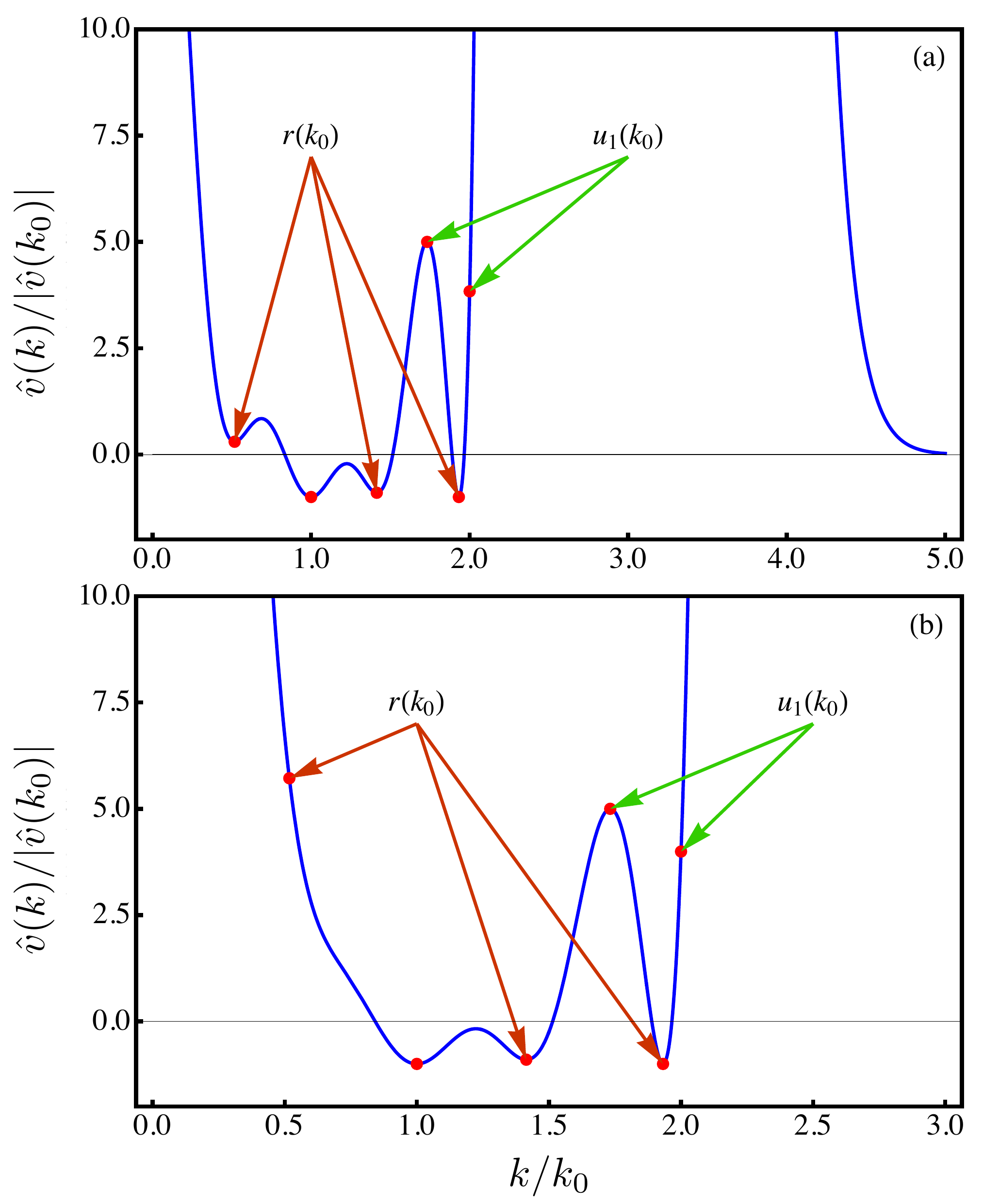}
	\caption{Examples of Lifshitz's potentials able to stabilize the QCP  as predicted from our variational development. 
	The full form of  $\hat{v}(k)$ in each case is obtained by imposing the following conditions: 
	(a) $n_{\mathrm{max}}=10$, $\hat{v}(0)=20$, $\hat{v}(\sqrt{3-\sqrt{2}}k_0)=0.3$, $\hat{v}(k_0)=-1$, $\hat{v}(\sqrt{2}k_0)=-0.9$, $\hat{v}(\sqrt{3}k_0)=5$ and  $\hat{v}(\sqrt{3+\sqrt{2}}k_0)=-1$ (in units of $|\hat{v}(k_0)|$) while $\sigma=1.2$; 
	and (b) $n_{\mathrm{max}}=8$,  $\hat{v}(0)=70$, $\hat{v}(k_0)=-1$, $\hat{v}(\sqrt{2}k_0)=-0.9$ and $\hat{v}(\sqrt{3}k_0)=5$ (in units of $|\hat{v}(k_0)|$) and $\sigma=1$. 
	In both panels the values of $\hat{v}(k)$ contributing to $r(k_0)$ and $u_1(k_0)$ are indicated.}
	\label{fig1.1}
\end{figure}

Our main goal is to study the possible stabilization of a twelve-fold symmetric dodecagonal quasicrystal phase (QCP) at zero temperature. We choose the nonlocal interaction potentials $v(r)$ of the Lifshitz's type~\cite{Barkan2014}, whose Fourier transform have the generic form
\begin{equation}
 \hat{v}(k)=\exp(-k^2\sigma^2)\sum_{n=0}^{n_{\mathrm{max}}}D_nk^{2n}.
\label{pot}
\end{equation}
The free parameters $D_n$'s and $\sigma$ can be tuned to guarantee a structure with several local minima at the desired wave vector, see Fig.\ref{fig1.1}. The high tuneability of this class of potential allowed to establish in the classical case that the stabilization of QCPs depends on the existence of a competition of different length scales\cite{Dotera2014,Barkan2014}

Our study of the ground state phase diagram is performed using the VMF approach~\cite{prestipino18,prestipino2018,likos01,likos01b,likos2008,glaser07}. Within the VMF approach the ground state wave function, $\psi(\{\mathbf{x}\})$, where $\{\mathbf{x}\}=\{\mathbf{x}_1,\dots \mathbf{x}_N \}$, is first chosen as the product of identical single particle wave functions
\begin{equation}
 \psi(\{\mathbf{x}\})=\prod_i\phi_0(\mathbf{x}_i).
 \label{fvar}
\end{equation}
Then, the normalized single particle wave function $\phi_0(\mathbf{x})$ is written as a Fourier expansion using a specific set of modes $\{c_j,\mathbf{k}_j\}$ in the form \cite{Mendoza-Coto2021a,Mendoza-Coto2021b} 
\begin{equation}
\phi_0(\mathbf{x})=\frac{c_0+\sum_{j\neq0}c_{j}\cos(\mathbf{k}_j\cdot\mathbf{x})/2}{\sqrt{A(c_0^2+\frac{1}{4}\sum_{j\neq0}c_{j}^2)}},
\label{anz}
\end{equation}
where $A$ is the area of the system. The set of Fourier modes considered in $\phi_0(\mathbf{x})$ defines the modulated pattern of the solution, as well as its symmetries.

\begin{table}[t!]
\begin{tabular}{@{}|p{.1\textwidth}|p{.2\textwidth}|c|@{}}
\cline{1-3}
    \textbf{Pattern} & 
    \textbf{Basis vector} $\mathbf{k}_{0,j}$ & 
    \textbf{Index range} \\ 
\cline{1-3}
    TCP     &  
    $k_0(\cos\frac{2\pi j}{6},\sin\frac{2\pi j}{6})$ &  
    $j=0,1$ \\
\cline{1-3}
    SqCP    &  
    $k_0(\cos\frac{2\pi j}{4},\sin\frac{2\pi j}{4})$   &  
    $j=0,1$ \\
\cline{1-3}
    Stripes &  
    $k_0(1,0)$ &  
    \\
\cline{1-3}
    12-QCP &  
    $k_0(\cos\frac{2\pi j}{12},\sin\frac{2\pi j}{12})$ &
    $j=0, 1,..., 5$ \\
\cline{1-3}
\end{tabular}
\caption{Modulated patterns studied in this work and the corresponding basis vectors used to generate the solutions $\phi(\mathbf{x})$. }
\label{tpatterns}
\end{table}

In this work we consider several ansatze for $\phi_0(\mathbf{x})$ to minimize the Hamiltonian Eq.~\ref{Hdim}. 
More precisely, we take into account the homogeneous solution,
a generic dodecagonal quasicrystal, and all possible periodic and symmetric patterns in two dimensions.
For the special case of the homogeneous solution all $c_j$'s vanish except for $c_0$.
In the case of modulated patterns, symmetry considerations allow us to significantly reduce the number of independent Fourier amplitudes ${c_j}$ by setting equal the Fourier modes corresponding to equivalent wave vectors $\mathbf{k}_j$.
The set $\{\mathbf{k}_{j}\}$ is constructed considering all possible combinations of a predetermined finite number of wave vectors, taken as the basis of $\{\mathbf{k}_j\}$, which are specific for each modulated solution.
In this way, by fixing the number of vectors of the basis that can be combined to form $\{\mathbf{k}_{j}\}$, we can establish the order of the ansatz.
The anzatze are summarized in Table~\ref{tpatterns}.

In all considered solutions the quantity $k_0$ represents the modulus of the wave vectors of the basis, which give us the scale of the modulation length of the respective pattern~\cite{Toner1981,Darci2007}. Fourier amplitudes $c_j$'s and $k_0$ are the variational parameters in the energy minimization process for each kind of solution.   

The total energy per particle $E/N$ of the model then reads
\begin{eqnarray}
 \frac{E}{N}&=&\langle\psi\vert\hat{T}_1\vert\psi\rangle+\frac{(N-1)}{2}\langle\psi\vert\hat{V}_{12}\vert\psi\rangle\\ \nonumber
 &=&\frac{1}{4}\frac{\sum_{j\neq0}c_j^2k_j^2}{(1+\frac{1}{2}\sum_{j\neq0}c_{j}^2)}
 \\&+& \rho\lambda^2 U\frac{\left(\hat{v}(0)b_0^2+\sum_{j\neq0}\hat{v}(k_j)b_j^2/2\right)}{2(1+\frac{1}{2}\sum_{j\neq0}c_{j}^2)^2}.
 \label{EVq}
\end{eqnarray}
In the above expression, we used the fact that $\mathbf{k}_j$ and $-\mathbf{k}_j$ have the same Fourier amplitude $c_j$. 
In addition, the coefficients $b_j$'s are defined as the Fourier amplitudes of $(\sum_{j=0}c_{j}\cos(\mathbf{k}_j\cdot\mathbf{x}))^2$, in this way the coefficients $b_j$'s can be written  in terms of $c_j$ using the relation  
$(\sum_{j=0}c_{j}\cos(\mathbf{k}_j\cdot\mathbf{x}))^2=\sum_{j=0}b_{j}\cos(\mathbf{k}_j\cdot\mathbf{x})$.
This energy is then compared to the energy of the homogeneous superfluid solution $\epsilon_{sf}=E_{sf}/N=\rho\lambda^2 U\hat{v}(0)/2$ ~\cite{macri13}.

\section{Single mode approximation} \label{singlemode}

We now identify the conditions to be satisfied by $\hat{v}(k)$ to stabilize the QCP over all other symmetric and periodic possible phases in two dimensions. 
We begin by performing the a simple analysis of the necessary conditions to guarantee that the QCP has a lower energy than the six-fold symmetric TCP.
The TCP configuration is selected as a benchmark for comparison with the QCP, since the triangular lattice corresponds to the optimal packing arrangement in two dimensions~\cite{cinti14,prestipino14,kroiss19}.

Within the single mode approximation, i.e. fixing all $c_j$'s to zero except for $c_1$, the energy per particle defined in Eq.\ref{EVq} of the TCP is given by
\begin{eqnarray}
\nonumber
  \epsilon_{t}&=&\frac{1}{4}\frac{3c_1^2k_0^2}{(1+\frac{3}{2}c_{1}^2)}+\frac{\rho\lambda^2U\hat{v}(0)}{2}\\
  &+&\rho\lambda^2U\frac{\left(3c_1^2(2+c_1)^2\hat{v}(k_0)+u_1(k_0)c_1^4\right)}{4(1+\frac{1}{2}3c_{1}^2)^2},
 \end{eqnarray}
where $u_1(k)=\left(3/4\hat{v}(2k) + 3 \hat{v}(\sqrt{3}k)\right)$.     
Modulated phases can usually be stabilized if the absolute minimum of $\hat{v}(k)$ is lower than zero and it simultaneously occurs at some finite wave vector modulus  $k_m$~\cite{likos01b}. In general, the variational treatment of the modulation wave vector $k_0$ produces non-trivial results which depend on the detailed form of $\hat{v}(k)$, even within the single mode approximation. 
In principle one might expect $k_0$ to be close to $k_m$ if the minimum of $\hat{v}(k)$ at $k_m$ is strong enough and we are close to the boundary between the homogeneous and the modulated phases, where we expect $c_1\ll1$.
Nevertheless, establishing rigorous conditions for $\hat{v}(k)$, under which the single mode is in fact a good approximation, is a very difficult task.

With the aim of identifying general ingredients, independent of the actual form of our $\hat{v}(k)$,
we begin by considering a single mode analysis in which $k_0$ is fixed to $k_m$.
In a subsequent step, we will perform a full variational analysis for specific cases of $\hat{v}(k)$ considering solutions with many modes to verify to which extent the conclusions about the stability of the QCP, from this simplified study, remain valid in the general case.    

Without loss of generality we take advantage of the fact that, in the appropriate units, the position of the main minimum of $\hat{v}(k)$ can be located at $k_0=1$, and its value set to $\hat{v}(k_0)=-1$. 

The other relevant phase in our simplified single mode analysis corresponds to the dodecagonal QCP. Considering Eq.~(\ref{EVq}) and the proposed ansatz for this phase, it is straightforward to conclude that the energy per particle for the QCP is given by  
\begin{eqnarray}
\nonumber
  \epsilon_\text{QCP}&=&\frac{1}{4}\frac{6c_1^2k_0^2}{(1+\frac{6}{2}c_{1}^2)}+\frac{\rho\lambda^2U\hat{v}(0)}{2}\\
  &+&\rho\lambda^2U\frac{\left(6c_1^2(2+c_1)^2\hat{v}(k_0)+u_2(k_0)c_1^4\right)}{4(1+\frac{1}{2}6c_{1}^2)^2},
 \end{eqnarray}
where $u_2(k)=2u_1(k)+r(k)$ and $r(k)=6(\hat{v}(\sqrt{2}k) + \hat{v}(\sqrt{2-\sqrt{3}} k) + \hat{v}(\sqrt{2 + \sqrt{3}}k))$. 

\begin{figure}[t!]
	\includegraphics[width=\linewidth]{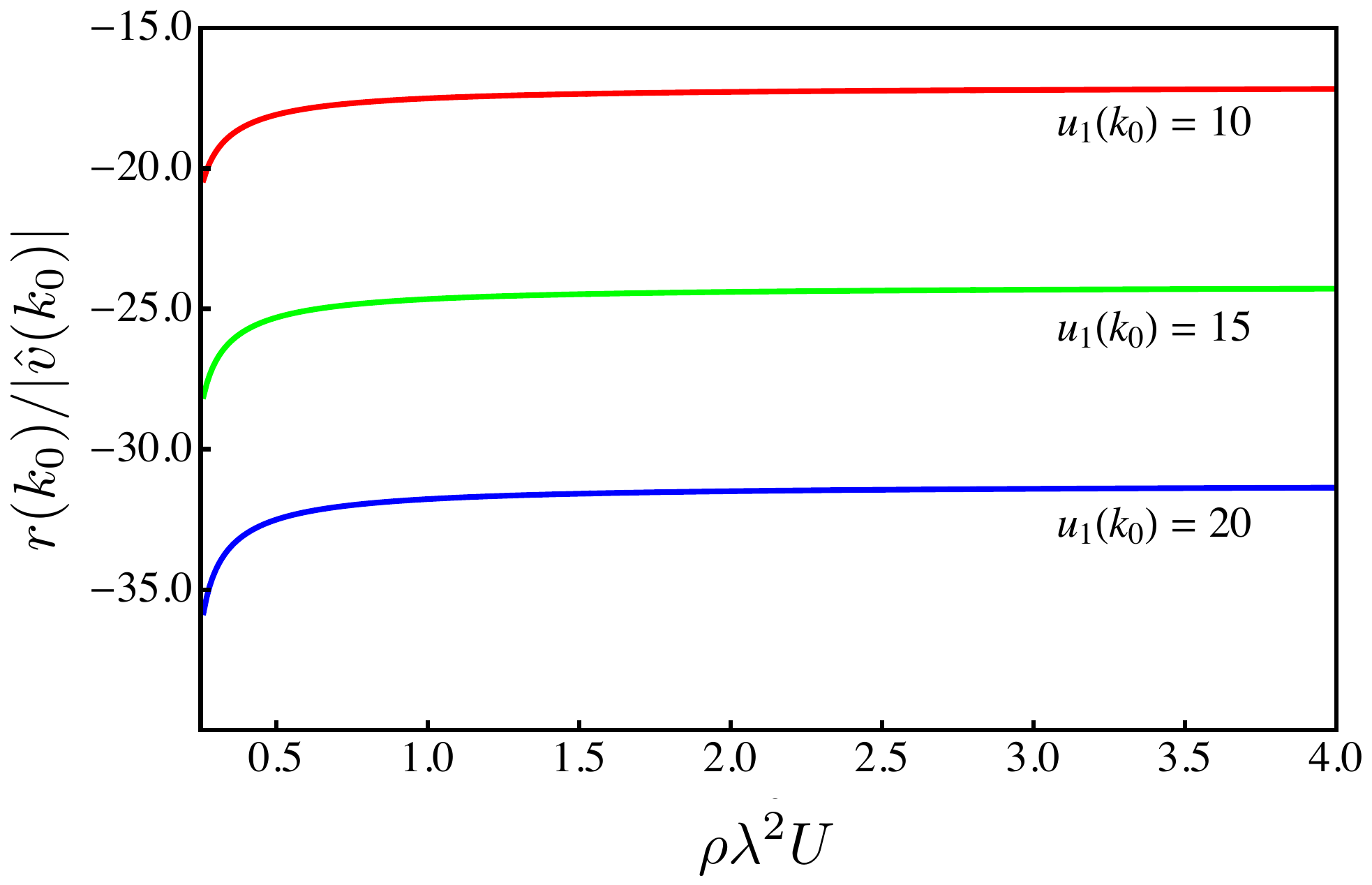}
	\caption{Boundary between the dodecagonal cluster quasicrystal phase and the triangular cluster crystal in the $r(k_0)$-$\rho\lambda^2U$ plane for $u_1(k_0)=10$, $15$ and $20$, in units of $|\hat{v}(k_0)|$. For each case the regions below the curve corresponds to those cases in which the dodecagonal solution has lower energy than the triangular lattice.}
	\label{fig1}
\end{figure}

It can be observed that the influence of $\hat{v}(k)$ in the energy per particle of the two relevant phases is encoded in two independent parameters, $r(k_0)$ and $u_1(k_0)$. 
Now we can compare the energy per particle of the TCP and QCP after minimizing with respect to the variable $c_1$, fixing the value of $u_1(k_0)$ and varying the parameters $\rho\lambda^2U$ and $r(k_0)$. 
We observe in Fig.~\ref{fig1} that, for large enough $\rho\lambda^2U$ values, the QCP becomes stable if $r(k_0)$ is low enough. 
This result confirms our initial expectation that if we decrease the value of the quartic coefficient for the QCP ($u_2(k_0)$), while maintaining $u_1(k_0)$ at a moderate to high value, the relative stability of the QCP is increased.

We can infer now that a sufficient condition for $\hat{v}(k)$ to stabilize the QCP will be to have low enough local minima at $\sqrt{3-\sqrt{2}}k_0$, $k_0$, $\sqrt{2}k_0$ and $\sqrt{3+\sqrt{2}}k_0$, to decrease $r(k_0)$; and to have local maxima or at least moderate values, at $\sqrt{3}k_0$ and $2k_0$, in order to obtain ``high'' values of $u_1(k_0)$. 
For future reference we will denote the values of $\hat{v}(k)$ at these particular points as characteristic values of the pair interaction potential. 
Additionally it is observed that, given the form of the Lipshitz potential, a high value of $\hat{v}(2k_0)$ is already guaranteed by the sequence of maxima and minima of the potential.
It can be concluded that a potential with eleven $D_n$'s independent coefficients ($n_{\mathrm{max}}=10$) is needed in order to build a $\hat{v}(k)$ for which all the characteristic values can be fine tuned.  

This discussion raises the question about what would be the actual minimum value of  $n_{\mathrm{max}}$ to maintain the stability of the QCP once we abandon the single mode approximation. 
We have studied this problem numerically by means of a many modes variational treatment of Eq.~(\ref{EVq}), considering the QCP and all other periodic solutions in two dimensions.
Our results indicate that the secondary minimum at $\sqrt{2-\sqrt{3}}k_0$ is in fact not a necessary ingredient beyond the single mode analysis. 
To the best of our knowledge, the simplest Lifshitz model stabilizing the QCP is the one with $n_{\mathrm{max}}=8$.     

\section{Phase Diagram characterization}\label{phasediagram}
\label{Pdiag}

\begin{figure}[t!]
	\includegraphics[width=\linewidth]{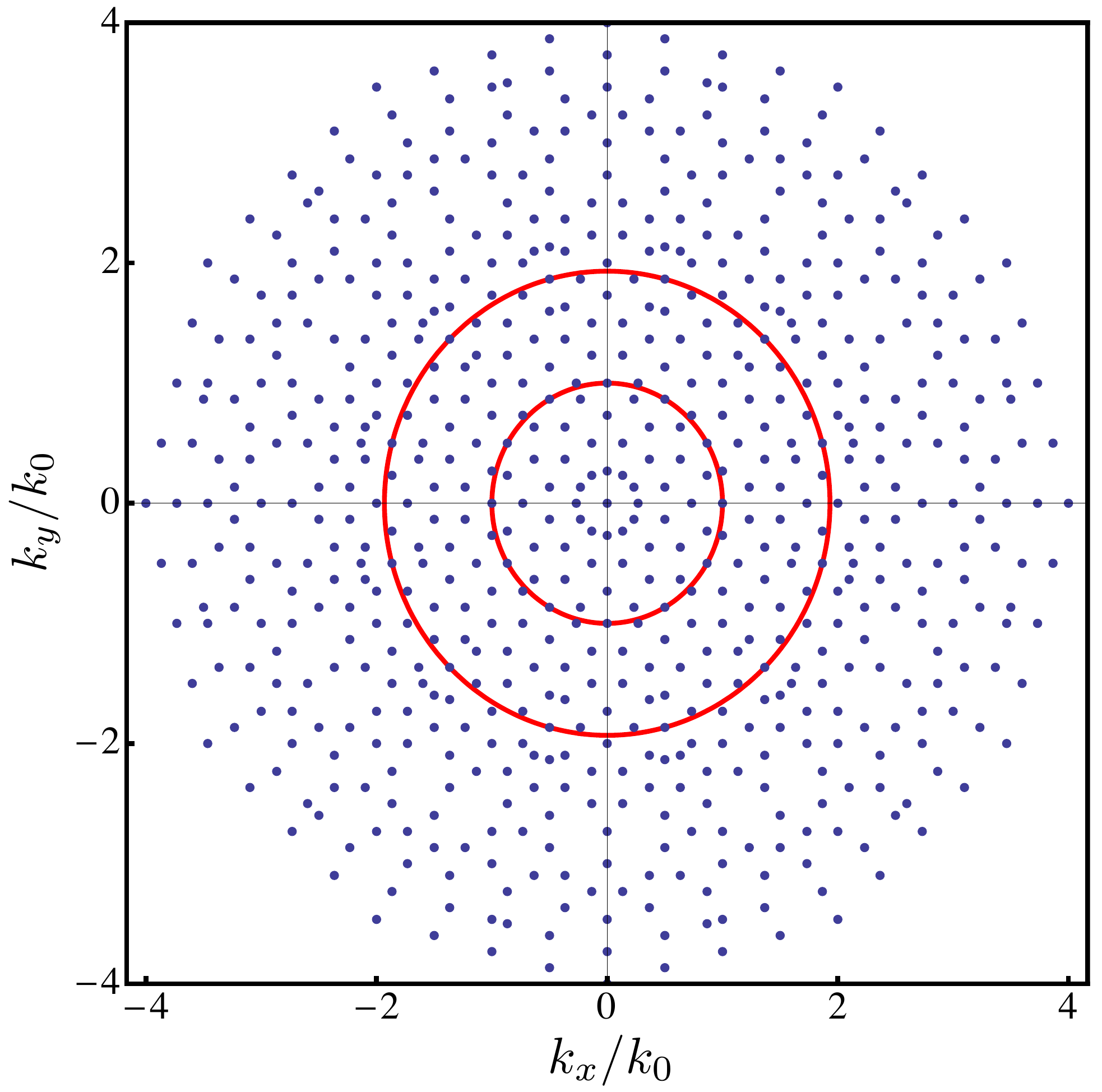}
	\caption{Wave vectors considered in the construction of the fourth order dodecagonal solution in units of the basis wave vector $k_0$. 
	Red circumferences 
	correspond to the two degenerate minima of the Fourier transform of the potential $\hat{v}(k)$. 
		}
	\label{fig2}
\end{figure}

Before presenting the results of the  numerical study, we provide some details about the construction of the solutions in the case of the many modes numerical analysis. Due to the aperiodic nature of the dodecagonal pattern, the set of wave vectors corresponding to the Fourier modes expansion of this solution rapidly increase when even a moderate number of possible combinations of wave vectors of the basis is considered. In the numerical analysis we consider a fourth order ansatz for the QCP solution, which means that all vectors resulting from the combinations of four vectors of the basis and the null vector will be considered. The resulting set of wave vectors considered in the construction of this solution is shown in Fig.~\ref{fig2}. This selection implies that the QCP solution has $37$ independent Fourier amplitudes ($c$'s in Eq.~(\ref{anz})), which are then considered as variational parameters, jointly with the scale of the main wave vector $k_0$. The number of variational Fourier amplitudes for each periodic solution is given by: $34$ for the TCP, $33$ for the SqCP and $10$ for the stripes solution. 

\begin{figure}[t!]
	\includegraphics[width=\linewidth]{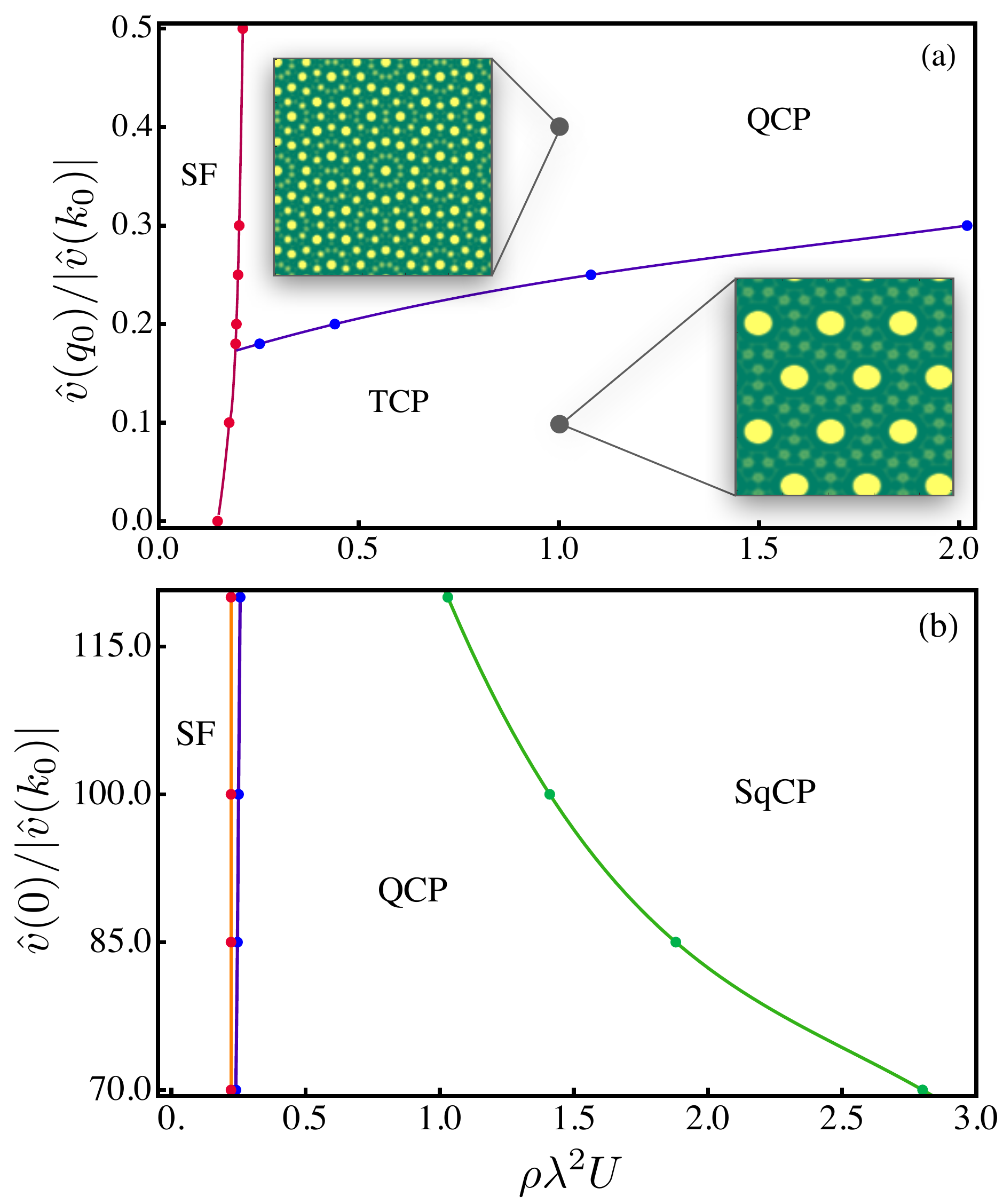}
	\caption{Ground state phase diagrams in the $\rho\lambda^2U$ versus $\hat{v}(q_0)$ plane for the two families of potentials given by Eq.~(\ref{pot}), using $n_{\mathrm{max}}=10$ (a) and $n_{\mathrm{max}}=8$ (b). The parameters of the potential are: (a) $\sigma=1.2$, $q_0=\sqrt{2-\sqrt{3}}k_0$, and subjected to the constraints $\hat{v}(0)=20$, $\hat{v}(\sqrt{2}k_0)=-0.9$, $\hat{v}(\sqrt{3}k_0)=5$, $\hat{v}(\sqrt{3+\sqrt{2}}k_0)=-1$, in units of $|\hat{v}(k_0)|$. Insets display two configurations referring to the TCP ($\rho\lambda^2 U=1$ $\hat{v}(q_0)/|\hat{v}(k_0)|=0.1$) and QCP phase ($\rho\lambda^2 U=1$ $\hat{v}(q_0)/|\hat{v}(k_0)|=0.4$), respectively. Configurations have been obtained solving numerically the Gross-Pitaevskii equation (see Section~\ref{supersolid}). (b) $\sigma=1$, $q_0=0$,  while subjected to the constraints $\hat{v}(\sqrt{2}k_0)=-0.9$, $\hat{v}(\sqrt{3}k_0)=5.0$, in units of $|\hat{v}(k_0)|$. }
	\label{fig12}
\end{figure}

In Fig.~\ref{fig12}a the results of the minimization are shown for a class of potentials given by the following set of $n_{\mathrm{max}}=10$ characteristic values: $\hat{v}(0)=20$, $\hat{v}(\sqrt{3}k_0)=-0.9$, $\hat{v}(\sqrt{2}k_0)=5$ and $\sigma=1.2$; $\hat{v}(\sqrt{2-\sqrt{3}}k_0)\in [0,1/2]$ is left as a free parameter.

It can be observed that, for low values of $\hat{v}(\sqrt{2-\sqrt{3}}k_0)$, as the product $\rho\lambda^2U$ is increased there is a transition from the homogenous superfluid phase (SF) to the TCP. This means that for such pair interaction potentials the TCP is favoured over the QCP (see Fig.~\ref{fig12}a). However, as this quantity is increased up to some critical value of $\hat{v}(\sqrt{2-\sqrt{3}}k_0)$, a region of stability of the QCP at intermediate values of $\rho\lambda^2U$ is developed.

Interestingly enough, the previous results show that large values of $\hat{v}(\sqrt{3-\sqrt{2}}k_0)$ do not eliminate the stability of the QCP. This fact suggests that a low value of $\hat{v}(\sqrt{2+\sqrt{3}}k_0)$ is not actually a necessary condition for the stability of the QCP beyond the single mode approximation. It implies that we can further reduce the order of the polynomial considered in the definition of Eq.~(\ref{pot}). This is related to the fact that the distribution of modes forming the QCP solution is much denser than the one we have for a periodic lattice. Consequently, there is a greater number of modes with wave vectors close to the optimal ones, which increase significantly the stability of the solution.

Considering the discussion above, we  analyze now the case of a simpler model with $n_{\mathrm{max}}=8$. As in the previous case all free parameters $D_n$'s will be determined from the characteristic values of the potential: $\hat{v}(\sqrt{2}k_0)=-0.9$, $\hat{v}(\sqrt{3}k_0)=5.0$ and $\sigma=1$, while $\hat{v}(0)\in [70,120]$ was taken as the free parameter. The results of the minimization process are shown in Fig.~\ref{fig12}b.

As in Fig~\ref{fig12}a, the QCP corresponds to the most stable phase over a wide $\rho\lambda^2U$ region. However, it can be noticed that there is not a direct transition from the homogeneous SF phase to the QCP. Instead, there is a narrow region between these two phases in which the TCP is the most stable phase. Additionally, for large enough values of $\rho\lambda^2U$, there is always a transition to the SqCP.

\section{Stability of the QCP}\label{stability}

To analyze the stability of the QCP with respect to small perturbations of the form of the potential, we 
consider the energy per particle given in Eq.~(\ref{EVq}) for arbitrary ansatze of the ground state wave function, which can be recast as 
\begin{equation}
 \frac{E}{N}=\langle \phi\vert \ \hat{T} \vert\phi\rangle+\frac{u}{2}A\langle \phi^2\vert v(x)\vert\phi^2\rangle,
\end{equation}
where $\vert\phi\rangle$ stands for the single state wave function, defined in Eq. (\ref{anz}). 
The first and second terms in the rhs of Eq.~(\ref{anz}) correspond to the average kinetic and potential energy per particle, while the parameter $u$ is a shorthand for the product $\rho\lambda^2U$.  

We assume that for a pair interaction potential $\hat{v}_0(k)$, the QCP is the lowest energy state per particle in the interval $u\in(u_1,u_2)$. As a consequence, in this interval, $\epsilon_q(u)<\epsilon_2(u)$, where $\epsilon_q(u)$ and $\epsilon_2(u)$ represent the energy per particle of the QCP and that of a secondary phase, respectively.

In the presence of a small perturbation $\delta\hat{v}(k)$ of the pair interaction potential, the energy per particle of the QCP yields
\begin{eqnarray}
 \epsilon_q&=&\epsilon_{q,0}(\{c_j\},k_0)+u\frac{\left(\delta\hat{v}(0)b_0^2+\sum_{j\neq0}\delta\hat{v}(k_j)b_j^2/2\right)}{2(1+\frac{1}{2}\sum_{j\neq0}c_{j}^2)^2},\nonumber\\
\label{eqc}
\end{eqnarray}
where $\epsilon_{q,0}(\{c_j\},k_0)$ stands for the energy function of the QCP in the absence of perturbation, $\{c_j\}$ is the set of Fourier coefficients of the QCP solution of the full problem and $k_0$ the corresponding modulation wave vector.
We then expand the energy per particle $\epsilon_q$ of Eq.~(\ref{eqc}) up to first order in $\delta\hat{v}(k)$, considering first that the $\{c_j\}$ and $k_0$ corresponding to the optimal solution of the perturbed problem can be also expanded in powers of $\delta \hat{v}(k)$. 
After some algebra we obtain
\begin{eqnarray}
\nonumber
\epsilon_q&=&\epsilon_{q,0}(\{c_{j,0}\},k_{0,0})\\ 
&+&u\frac{\left(\delta\hat{v}(0)b_{0,0}^2+\sum_{j\neq0}\delta\hat{v}(k_{j,0})b_{j,0}^2/2\right)}{2(1+\frac{1}{2}\sum_{j\neq0}c_{j,0}^2)^2},
\label{eqpert}
\end{eqnarray}
where $\{c_{j,0}\}$ and $k_{0,0}$, represent the optimal Fourier amplitudes and modulation wave vector of the unperturbed problem, respectively.
Here, we have already taken into account that $\partial_{c_j}\epsilon_{q,0}(\{c_{j,0}\},k_{0,0})=0$ and $\partial_{k_0}\epsilon_{q,0}(\{c_{j,0}\},k_{0,0})=0$, as well as the fact that the lowest order corrections to ${c_j}$ and $k_0$, corresponding to the optimal solution of the perturbed problem, are linear in $\delta\hat{v}(k)$.

The unperturbed problem could admit, in principle, more than one solution of the same kind. Considering that our analysis relies on the stability of the solutions of the original problem against perturbations, a required condition is the existence of a finite energy gap between the energy corresponding to the optimal solution and those corresponding to any other solutions of the same kind. If such a gap exists, then it is always possible to use perturbation theory around the optimal solution of the unperturbed problem for small enough $\delta \hat{v}(k)$.
The result obtained for the energy correction of the optimal QCP solution is also valid for other phases, since at any point we have taken advantage of the particular form of the QCP solution. 
The corrected energy per particle as a function of the parameter $u$ is
\begin{eqnarray}
\nonumber
 \epsilon_q(u)&=&\epsilon_{q}^0(u)+\frac{u}{2}A\langle\phi_{q,0}^2\vert \delta v(x)\vert\phi_{q,0}^2\rangle,\\
 \epsilon_2(u)&=&\epsilon_{2}^0(u)+\frac{u}{2}A\langle\phi_{2,0}^2\vert \delta v(x)\vert\phi_{2,0}^2\rangle.
\end{eqnarray}

Now it is possible to compute, up to first order in $\delta{v}(x)$, the solution of the equation $\epsilon_q(u)=\epsilon_2(u)$, which gives the location of the phase boundary after perturbing the pair interaction potential. 
Considering that at $u=u_1$, $\epsilon_{q}^0(u_1)=\epsilon_{2}^0(u_1)$ holds, 

we conclude that up to first order in $\delta v(x)$ the correction to the phase boundary position reads 
\begin{equation}
 \delta u=\frac{u_1}{2}\frac{\left(A\langle\phi_{2,0}^2\vert \delta v(x)\vert\phi_{2,0}^2\rangle-A\langle\phi_{q,0}^2\vert \delta v(x)\vert\phi_{q,0}^2\rangle\right)}{\left(\partial_u\epsilon_{q}^0(u_1)-\partial_u\epsilon_{2}^0(u_1)\right)}.
\end{equation}

This result shows that, for certain $\hat{v}(k)$, the QCP corresponds to the ground state of the system in a certain finite region of the parameter $\rho\lambda^2 U$, in which arbitrary small enough perturbations of $\hat{v}(k)$ will not destroy the stability of the QCP.

\section{Elementary excitations}\label{excitations}

Now we focus on the excitation properties of the homogeneous phase in the presence of potentials stabilizing the QCP. Within the Bogoliubov theory, the low energy excitation spectrum within the superfluid phase can be written as  \cite{macri13,ancilotto13}
\begin{equation}
 \epsilon(k,\rho\lambda^2 U)=\sqrt{\frac{k^2}{2}\left(\frac{k^2}{2}+2\rho\lambda^2 U\hat{v}(k) \right)}.
\end{equation}
In Fig.~\ref{fig1t}d the excitation spectrum for several values of $\rho\lambda^2 U$ is shown. For comparison, a plot of the $\hat{v}(k)$ used in these calculations is also included. 

The sequence of local minima in $\epsilon(k)$, for intermediate values of $\rho \lambda^2 U$, closely reproduces the sequence of local minima in $\hat{v}(k)$. At large enough values of $\rho \lambda^2 U$, the excitation spectrum develops several roton minima corresponding to the various minima of $\hat{v}(k)$. Interestingly, the dominant roton minimum position changes from $\sqrt{2-\sqrt{3}}k_0$ to $k_0$ as the density is increased, due to the competition between the kinetic energy term and the pair interaction potential in $\epsilon(k)$. In this way, it is precisely the roton minimum at $k=k_0$ the one responsible for the destabilization of the homogeneous phase (see Fig.~\ref{fig1t}). 

Another quantity of interest which is accessible from the Bogoliubov theory is the so-called normalized condensate depletion $f_{nc}$ characterizing the non-condensed fraction of the system 
\begin{equation}
f_{nc}=\frac{1}{\rho}\int\frac{d^2k}{(2\pi)^2}\frac{1}{2}\left[\frac{\frac{k^2}{2}+\rho_0 U\hat{v}(k)}{\epsilon(k,\rho_0 U)}-1\right],
 \label{eq:cond_frac}
\end{equation}
where $\rho_0$ represent the particle density of the condensate. The non-condensed fraction can be determined self-consistently from the relation $\rho_0=\rho(1-f_{nc})$, a condition that guarantees the proper normalization of $f_{nc}$, i.e. $0\leq f_{nc}\leq1$.

In Fig.~\ref{fig1t}e, a phase diagram $\rho\lambda^2$ versus $U$ is presented describing the Bose-Einstein condensation in the homogeneous state for the model given by Eq.~(\ref{pot}), using the same set of characteristic values considered in Fig.~\ref{fig1.1}a. It can be observed a crossover of $f_{nc}$ from low to high values as $U$ is increased. We have highlighted with a green curve the boundary of the region of high non-condensed fraction ($f_{nc}>0.9$). This boundary is not monotonic, revealing a nontrivial interplay between the density and the potential strength. 


\section{Supersolidity within the QCP}\label{supersolid}

The supersolid phase is a state of matter that breaks both continuous translational and global $U(1)$ symmetries, exhibiting simultaneously a crystalline order and off-diagonal long range order~\cite{Boninsegni2012,cinti14}. 
Many efforts from the theoretical perspective \cite{Chester1970,Leggett1970,Boninsegni2012,Cinti2010,henkel10,cinti14,Li2013,Liao2018} as well as several low temperature experiments with dipolar quantum gases \cite{PhysRevLett.123.050402,Tanzi_2019,PhysRevLett.122.130405,PhysRevX.9.011051,Chomaz2019,Norcia_2021} have been realized in recent years to understand the properties and the existence of these density modulated superfluid systems.

In order to analyze if a supersolid-like phase could be stabilized within the QCP, we consider two parameters quantifying both superfluid and quasi-crystalline order. We employ the Leggett’s criterion \cite{Leggett1970,zhang19} which allows to compute an upper bound for the superfluid fraction as
\begin{equation}
    f_s=\mathrm{Min}_{\theta}\left[\int \frac{d^2x}{A}\frac{1}{\int_{0}^L\frac{dx}{L}\rho(x',y')^{-1}}\right],
    \label{supf}
\end{equation}
where the function $\rho(x,y)=A|\phi_0(x,y)|^2$ and $A$ and $L$ stand for the area and linear dimension of the system, respectively.
 
In this equation, we should take the minimum with respect to all possible directions defined by the angle $\theta$, taking $x'=x\cos{\theta}-y\sin{\theta}$ and $y'=x\sin{\theta}+y\cos{\theta}$. Instead of proceeding directly with the numerical calculation of $f_s$ it is convenient first to discuss some mathematical properties of the quantity defined in Eq.~(\ref{supf}), which can lead to a simplification of the numerical evaluation of $f_s$.

Let us begin analyzing the quantity $\int dx/L \rho^{-1}(x',y')$ in the limit $L\rightarrow\infty$, which is in principle a function of $y$ and $\theta$. 
If $\phi_0(\mathbf{x})$ is a periodic or quasi-periodic function, then $\rho^{-1}(\mathbf{x})$ will have the same symmetry properties of $\phi_0(\mathbf{x})$, and consequently the same full set of Fourier modes can be used in general to expand $\rho(\mathbf{x})$ and $\rho^{-1}(\mathbf{x})$. 

Therefore, without lost of generality, $\rho^{-1}(\mathbf{x})$ can be written as
\begin{eqnarray}
\nonumber
    \rho^{-1}(x',y')&=&d_0+\sum_{i\neq0}d_i\cos((k_{ix}\cos(\theta)+k_{iy}\sin(\theta))x\\
    &+&(k_{iy}\cos(\theta)-k_{ix}\sin(\theta))y),
    \label{eqrho1}
\end{eqnarray}
where $\{k_{ix},k_{iy}\}$ represent the cartesian components of $\mathbf{k}_i$ and the $d_i$'s represent the Fourier amplitudes of $\rho^{-1}(x,y)$, defined in the usual way. Proceeding with the formal integration along the $x$-variable, and taking $L\rightarrow\infty$, we find 
\begin{eqnarray}
\nonumber
&&\int_{0}^L\frac{dx}{L}\rho^{-1}(x',y')=d_0+
\sum_{i\neq0}d_i 
\cos((k_{iy}\cos(\theta)\\
&-&k_{ix}\sin(\theta))y)
\delta(k_{ix}\cos(\theta)+k_{iy}\sin(\theta),0),
\label{Ftx}
\end{eqnarray}
where $\delta(a,b)$ stands for the Kronecker delta function. 
This result implies that, unless $\theta$ is selected to be one of the possible discrete values for which  $k_{ix}\cos(\theta)+k_{iy}\sin(\theta)=0$, the result of the integration is a constant equal to $d_0$. 

Now we can take advantage of the  Schwartz inequality, which allows  us to conclude directly that 
\begin{equation}
\int \frac{dy}{L}{\frac{1}{\int\frac{dx}{L}\rho^{-1}(x',y')}}\geq\frac{1}{\int \frac{dy}{L}\int\frac{dx}{L}\rho^{-1}(x',y')}    
\end{equation}
Considering then the form of Eq.(\ref{Ftx}), or even Eq.(\ref{eqrho1}), is straightforward to conclude that
\begin{equation}
\int \frac{dy}{L}{\frac{1}{\int\frac{dx}{L}\rho^{-1}(x',y')}}\geq\frac{1}{d_0}.     
\end{equation}
Since this inequality holds for all $\theta$ and only becomes an identity when $\theta$ corresponds to one of those values which makes zero the oscillatory dependence in $y$ of the r.h.s. of Eq.(\ref{Ftx}), we can conclude that 
\begin{equation}
\mathrm{Min}_\theta\left[\int \frac{dy}{L}{\frac{1}{\int\frac{dx}{L}\rho^{-1}(x',y')}}\right]=\frac{1}{d_0}    
\end{equation}
This means that the superfluid fraction given by Eq.(\ref{supf}) is
$f_s=\frac{1}{d_0}$.

We observed that for large enough $\rho \lambda^2 U$ numerical issues in the variational minimization process eventually produce spurious solutions with nodes, leading to a vanishing superfluid fraction. This effect is not present in the full numerical solution of Gross-Pitaevskii equation\cite {macri13} (see the following discussion).

The crystalline order of the modulated patterns can be characterized by using the so-called density contrast, defined as \cite{PhysRevLett.122.130405,PhysRevX.9.011051,Chomaz2019}
\begin{equation}
C=\frac{max\left[\rho\left(\mathbf{x}\right)\right]-min\left[\rho\left(\mathbf{x}\right)\right]}{max\left[\rho\left(\mathbf{x}\right)\right]+min\left[\rho\left(\mathbf{x}\right)\right]}. 
\end{equation}
For states without density modulations, the parameter $C$ vanishes. 
On the other hand, for strongly density modulated states, 
$C$ is close to unity. 

In the case of quasicrystalline patterns, the determination of the contrast has the inherent complication related to the fact that the maximum and minimum of the profile density are not well defined single values. 
Instead, in that case, there is a distribution of local maxima and minima over the system and, in general, the calculus of the absolute minimum and maximum of the profile density is a difficult task. 

In our case, given the form of the ansatz for the single particle wave function, the absolute maximum is located at the origin of the coordinate system. 
However, the determination of the absolute minimum of the density profile is far from trivial. 
Because of this, we adopt the simplifying criterion of taking the minimum of the density profile as that of the local minimum closest to the absolute maximum. 
Such a value can be determined numerically by a simple minimization procedure of $\rho(\mathbf{x})$.

\begin{figure}[t]
    \includegraphics[width=\linewidth]{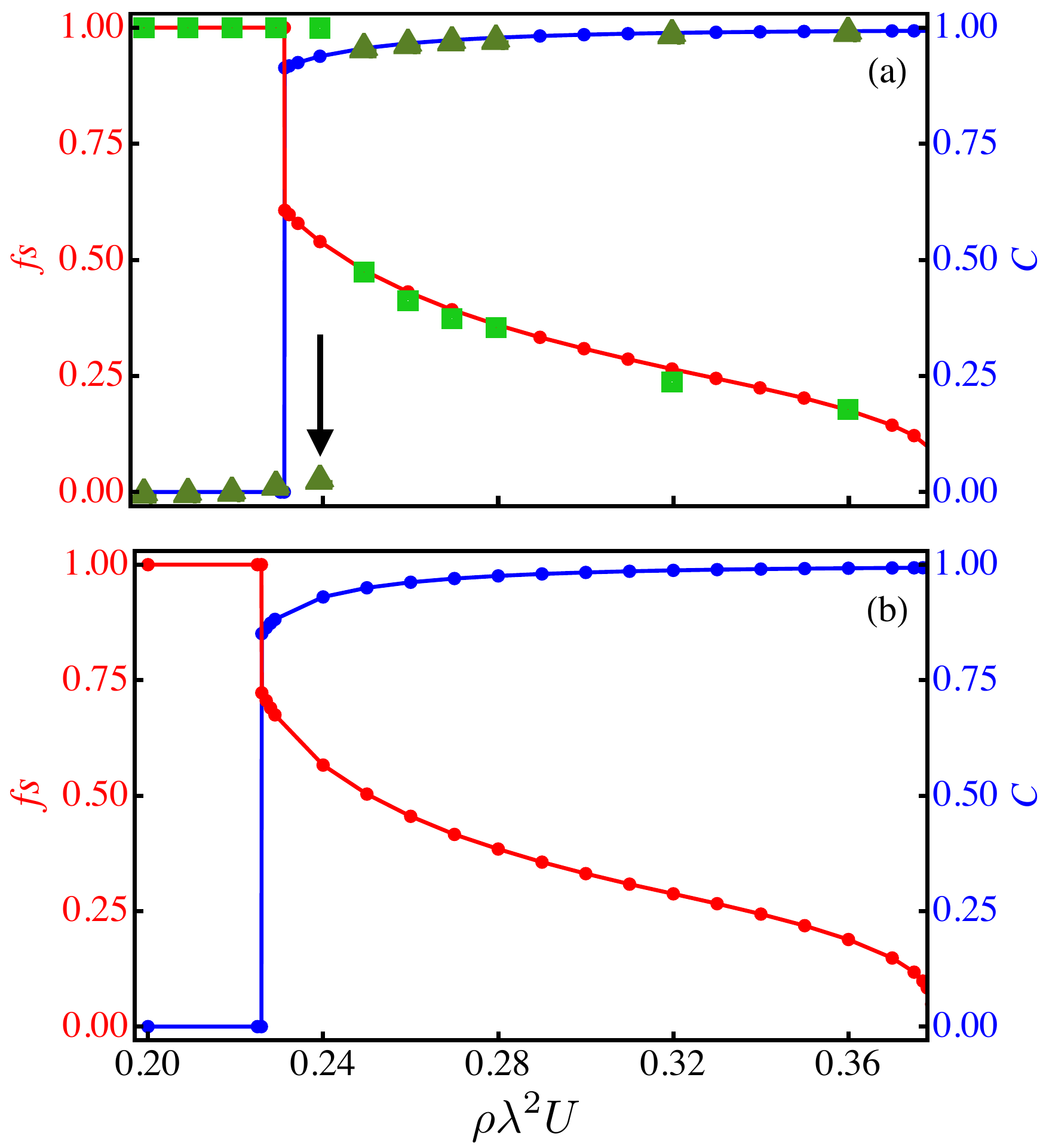}   
	\caption{Density contrast and superfluid fraction as function of $\rho \lambda^2 U$, setting $\hat{v}(\sqrt{3-\sqrt{2}}k_0)=5$ (a) and $\hat{v}(\sqrt{3-\sqrt{2}}k_0)=10$ (b). 
	The rest of the characteristic parameters of the potential are the same as in Fig~\ref{fig1.1}a. 
	Two different phases can be distinguished: the superfluid phase in which $C=0$ and $f_{s}=1$ and the supersolid-like QCP in which $C>0$ and $0<f_{s}<1$. In (a) squares and triangles in green correspond to the superfluid fraction and to the density contrast computed from the direct numerical solution of the Gross-Pitaevskii equation respectively. Additionally, the black arrow mark the transition from the superfluid to the supersolid like QCP also obtained from Gross-Pitaevskii simulations.
		}
	\label{fig9}
\end{figure} 
In Fig.~\ref{fig9} the behavior of both the superfluid fraction $f_{s}$ and the density contrast $C$ is presented for the model given by Eq.~(\ref{pot}) with $n_{\mathrm{max}}=10$, considering two different inputs for the characteristic value $\hat{v}(\sqrt{3-\sqrt{2}}k_0)$. The obtained results confirm in both cases a sequence of superfluid and supersolid-like QCP phases as $\rho \lambda^2 U$ is increased from low values. A discontinuous phase transition is clearly observed from the homogeneous SF phase to the supersolid-like QCP phase at $\rho\lambda^2 U\sim0.23$. As a test for the validity of the MF variational observations presented in Fig.~\ref{fig9}a we have computed $f_s$ and $C$ for this case from numerical simulations of the Gross-Pitaevskii equation. We observe that the first order transition is slightly shifted with respect to the variational value. After this transition region the agreement obtained between analytical and numerical simulation results is excellent giving in this way a strong validation of our variational study.  

\section{Conclusions}
\label{conclusions}

In the present work we analyzed under which conditions a cluster quasicrystal phase is self-stabilized in a $2$D system of interacting bosons at zero temperature. We used the VMF firstly to identify the necessary ingredients to stabilize a dodecagonal quasicrystal modulated pattern in a model interacting through a Lifshitz-type potential and, in a second stage, to systematically study the complete phase diagram of these models varying the form of the pairwise potential. 
Our numerical studies considered several ansatze for the modulated phases.
We obtained that, depending on the form of the pair interaction potential, the cluster QCP can be stabilized in a wide $\rho \lambda^2 U$ interval, ranging from the boundary of the homogeneous phase to the classical regime at large $\rho \lambda^2 U$ values. 

This scenario suggested the possibility of the existence of a supersolid-like QCP and a classical QCP in which the superfluid fraction is finite and zero, respectively. The stability of this QCP against small variations of $\hat{v}(k)$ was also confirmed. Additionally, we showed that, once the QCP is stable for a given $\hat{v}(k)$ in a certain region of the parameter $\rho \lambda^2 U$, small enough variations $\delta\hat{v}(k)$ in the pair interaction potential change smoothly the $\rho \lambda^2 U$ region of stability of the QCP.

The excitation properties within the homogeneous phase for models stabilizing the QCP were studied by monitoring the Bogoliubov spectrum for a wide interval of $\rho \lambda^2 U$. In general, it was observed a structure of local minima that closely follows the one observed for the pair interaction potential. For $\rho \lambda^2 U$ close to the limit of stability of the homogeneous phase, a dominant roton minimum is developed at the wave vector corresponding to the main minimum of the pair interaction potential $(k=1)$, signaling an instability towards the formation of modulated patterns with this characteristic wave vector. We found that the limit of Bogoliubov stability of the homogeneous phase is rather close to the actual boundary of the superfluid and homogeneous phases. 

We studied simultaneously the superfluid fraction $f_{s}$, estimated using a well established adaptation of Legget's criterion~\cite{Leggett1970,zhang19}, and the density contrast $C$. Our results suggest that, close to the superfluid phase boundary (see Fig.~\ref{fig9}), the QCP hosts a supersolid state in which simultaneously emerges quasicrystalline order and superfluidity~\cite{Pupillo2020}. The combination of a many-mode variational minimization and an accurate calculation of the Legget's superfluid fraction allowed us to distinguish also the phase boundary of this supersolid-like QCP.

Finally we notice that, although we focussed on a selected class of interactions, the methodology applied here is general and can be used with other types of potentials. Our results provide a solid basis for the search of physical systems where tuneable two-body interactions are capable to stabilize many-body quantum quasicrystal phases of the kind described in this work.

\section{Acknowledgements}
A.M.C. acknowledges financial support from Funda\c{c}\~ao de Amparo \`a Pesquisa de Santa Catarina, Brazil (Fapesc). A.M.C. acknowledges R. Cenci for helpful discussions. R. T. thanks the Physics Department of the Universidade Federal de Santa Catarina for full support. This study was financed in part by the Coordenação de Aperfeiçoamento de Pessoal de Nível Superior - Brasil (CAPES) - Finance Code 001. 
T.M. acknowledges CNPq for support through Bolsa de produtividade em
Pesquisa n.311079/2015-6. T.M and V.Z. are supported by
the Serrapilheira Institute (grant number Serra-1812-27802).
The numerical integration of the Gross-Pitaevskii equation was done with the help of the XMDS2 software~\cite{xmds2}.

\bibliographystyle{apsrev4-1}
\bibliography{clusters.bib}

\end{document}